\documentclass{elsart}
\begin{document}
\begin{frontmatter}

\title{{Comment on: Bell's inequalities\\ II: Logical loophole in their
interpretation}} \author{A. F. Kracklauer}

\maketitle

\address{Institut f\"ur Mathematik und Physik; \\Bauhaus Universit\"at, 99401
Weimar, BRD}

\thanks{P.F. 2040, 99401 Weimar, BRD; kracklau@fossi.uni-weimar.de}

\begin{abstract}

In a remarkably insightful pair of papers recently, Sica demonstrated that:
dichotomic data taken in any experiment that violates Bell's inequalities 
``cannot represent any data streams that could possibly exist or be imagined''
if it is to be consistent with the derivation of the inequalities.\cite{sica}
The present writer maintains, however, that corrections in the formulation of
Bell's analysis loosen  restrictions imposed by Bell inequalities. Moreover, it
is argued that the resolution proposed by Sica for the conflict arising from
the fact that real data does violate Bell inequalities, namely that the
functional form of the correlations considered by Bell must be amended, is
untenable on physical grounds. Finally, an alternate resolution is proposed.


\end{abstract}
\end{frontmatter}

Sica examined the arithmetic of dichotomic sequences, i.e., lists of \( \pm 1,
\) and their correlations. With elementary analysis he showed that all such
sequences yield among themselves correlations which always satisfy Bell
inequalities. This leads immediately to the conclusion that all data taken in
experiments that generate such sequences must as a tautology satisfy Bell
inequalities.\cite{sica} It is well known, however, that experiments testing
Bell's inequalities produce results that violate them. 

The conventional understanding is that violation results because the
inequalities were derived with motivation and argumentation based on the
hypothetical existence of hidden variables, which, when included in a theory 
underlying Quantum Mechanics (QM), might somehow render the interpretation free
of well known problems or preternatural aspects. As experiments violate the
inequalities which were ostensibly derived solely on the basis of local,
realistic `natural philosophy,' it is generally understood that in so far as
empirical truth does not support Bell inequalities, nature is not local and
(or) realistic. In particular it is taken that nature at a fundamental level is
nonlocal; and therefore, hidden variables, if any are to be found, will be also
nonlocal.\cite{bell} (I leave it to the proponents of this viewpoint to say
what this really means.)

At this point, note that Barut provided a counter example to Bell's Theorem
thereby showing that it is not always correct.\cite{Barut} What Barut did not
do was to analyse the extraction of the inequalities to reveal their
limitations. This can be done as follows.

Barut's fundamental point is, that by introducing continious hidden variables,
it is possible to calculate the QM spin correlation for the EPR(B) gedanken
experiment, \( -\cos (\theta ), \) on the basis of a fully realistic and local
model. To do so, he used the full form for the correlation of two random
variables, \( A \) and \( B \):

\begin{equation}
\label{rancor}
Cor(A,\, B)=\frac{<|AB|>-<A><B>}{\sqrt{<A^{2}><B^{2}>}}.
\end{equation}

When \( A \) and \( B \) are dichotomic random variables taking on the values
\(  \pm 1 \), then \( <A>=<B>=0 \),  \( <A^{2}>=<B^{2}>=1 \) and the extraction
of Bell inequalities goes through just as Bell and others presented it.
However, if \( A \) and \( B \) are random variables for which these conditions
do not hold, then an additional term changes \( N \) in Bell's four-setting
version inequality; i.,e.,

\begin{equation}
\label{BellIn}
|P(a,b)-P(a,b')|+|P(a',b')+P(a',b)|\leq N,
\end{equation}

by the amount: \( +\frac{2<A><B>}{\sqrt{<A^{2}><B^{2}>}} \). 

These considerations do not affect the case considered by Sica, which is
restricted to variables for which both averages are zero. 

In order to resolve the conflict between Bell inequalities---which are after
all, arithmetic tautologies---and empirical truth, Sica proposed amending the
functional form of the intersequence correlations. This, however, has a very,
very high price! The actual experiments carried out to date that test Bell
inequalities have not used the spin-variant of the EPR gedanken experiment, but
a parallel one employing polarised `photons.' Polarisation of electromagnetic
signals is a well understood phenomena. The correlations existing between
different states of polarisation, which differs from the above form only in
being: \( -\cos (2\phi ), \) where \( \phi  \) is the angle between
polarisation modes or polarisers, have been confirmed beyond any doubt.
Rejecting this verity seems out of the question.

Thus, the situation seems to be that a dichotomic process, generating data
correlated per an empirically verified form, violates an arithmetic identity!
In view of the above, however, an alternate way out of this dilemma is to take
it that the correlation, \( -\cos (2\phi ) \), does not pertain to dichotomic
data for which the averages are zero---a conclusion won elsewhere with other
arguments.\cite{nodico} This can come about in the following way. First, some
preliminaries.

The fundamental dispute which Bell was addressing is: can QM be so extended
that it turns out to be a statistical covering theory for an underlying
classical theory involving extra, heretofore hidden, variables? In this spirit,
therefore, let us take it that the signals generated in the optical version of
the EPR gedanken experiment are in fact classical electromagnetic fields, not
photons. The EPR source then can be seen as emitting a symmetric but
unpolarised signal in all directions. Thus, the geometric structure of electric
field that reaches the \( A \) and \( B \) detectors would be \( E_{A}=\cos
(\theta )\hat{x} \) \( \pm \sin (\theta )\hat{y} \), and  \( B \), \(
E_{B}=\sin (\theta +\phi )\hat{x}\mp \cos (\theta +\phi )\hat{y} \),
respectively, where \( \theta  \) is the instantaneous polarisation angle and
the ambiguous signs are to be chosen to account for the four channels. Also,
factors of the form \( \exp(i(\omega t+\delta (t))) \)where \( \delta (t) \) is a
random function of \( t \), have been suppressed as they will all drop out with
averaging.\cite{aaad} The probability of a detection, in the end necessarily a
photoelectron, is proportional to the square of these fields. Thus, for these
signals, \( <A,B>=<E^{2}_{A,B}>=1 \) and \( <A^{2},B^{2}>=1 \) so that in this
case Bell's inequality is to be satisfied with \( N=4. \)

 Because electrodynamics is linear at the field level and not the intensity
level where statistics enter via ``square-law'' detectors, calculating
coincident probabilities (or  any other `would-be' product of intensities)
actually requires calculating fourth order field correlations instead of the
direct product of intensities; e.g.,  
\begin{equation}
 P(\pm \pm)=\frac{<(E_{A}\cdot
E_{B})(E_{B}\cdot E_{A})>}{<E^{2}_{A}+E^{2}_{B}>}.
\end{equation}

For the signals considered above, the pairwise coincident probabilities are \(
P(++)=P(--)=\sin ^{2}(\phi )/2; \) \( P(+-)=P(-+)=\cos ^{2}(\phi )/2 \),
yielding the correlation: \( -\cos (2\phi ) \), the same result as given by
QM. Furthermore, as the rhs of Eq. \ref{BellIn} is observed to be \( \leq
2\sqrt{2}, \) the appropriately modified Bell inequality is fully respected.

From the vantage of this model, the statistics of the EPR experiment are simply
due to the geometrical interplay of polarisers and unpolarised radiation. 
Neither needle radiation nor otherwise bundled and directed emission of wave
packets; a.k.a. `photons' are needed. Basically, an atom has the structure of a
dipole---an electron whirling about a proton---and it is extremely unlikely
that dipole radiation could be consitently generated with such low entropy
structure.  In this model, this dipole radiation behaves as if from a dipole
antenna and spreads in all directions. The number of detections (photoelectron
pairs) at any given setting of the detectors is  simply in proportion to the
matched intensity in both arms of the fields entering the detectors. 
Detections unmatched by a correspondent in the opposite arm are thrown out of
the sample; in experiments this is effected by the coincidence circuitry.  No
collapse or other superluminal interaction is needed.  By way of contrast, in
the ``Copenhagen'' interpretation, the singlet state is considered
fundamentally unrealized until ``projected'' out or ``collapsed'' at a
detector.  Then, when the polarisers are not parallel, the projection occurs
stochastically but in proportion to the angular geometry.  In order to make
this imagery hold together, one then has to consider superluminal coordination
of the projection process; and no justification is offered for coincidence
circuitry except to exclude signals from spurious sources. (Indeed, the
use of coincidence circuitry even seems to  conflict with the implicit
logic of the derivation of Bell inequalities.) 

In Accord with the imagery of this model, all coincidence events are simply
coincidences of independent single events within the window set by the
circuitry.  On the other hand, the orthodox view holds that all single,
unmatched events result from a failure of one or the other detector to register
one memeber of a simultaniously `projected-out'photon pair, except for those
events caused by spurious background sources.  This distinction should lead to
a different dependence of the total observed count rate on the coincidence
window width when the source intensity is so low that only one pair (of
photons) can be in play at a time.  In the former case the observed count rate
would be expected to be simply  proportional to the window width, while the
orthodox image implies a certain independence of the window width as only one
pair at a time is available for detection, even when coherence length and
counter efficiency are taken into account.  Such an effect might empirically
differentiate these paradigms.

Some readers may be uncomfortable with these arguments having noticed that
there is nothing distinctly quantum mechanical about them, that is, there was
no need to introduce Planck's  constant. This in the midst of a dispute to
plumb the innate character of QM! There is, however, nothing here
contradictory; QM itself maintains that polarisation phenomena are classical.
QM enters the picture where and only where noncommutivity is in evidence
between conjugate variables \emph{iff} their classical correspondents do
commute. The creation and annihilation operators for photons of different
polarisation modes, \emph{do} commute; i.e., there is nothing QM in their
nature. Noncommutivity of nonorthogonal polarisation states classically
reflects the fact that the order with which a signal traverses polarisers
matters. If a linearly polarised signal passes first through a polariser making
an angle \( \theta _{1} \)with its polarisation vector and then through a
second polariser making an angle \( \theta _{2} \) with respect to the first
polariser, the intensity is reduced by \( \cos ^{2}(\theta _{1})\cos
^{2}(\theta _{2}), \) whereas in the reverse order it would be \( \cos
^{2}(\theta _{1}+\theta _{2})\cos ^{2}(\theta _{1}). \) These operations do not
commute but this has nothing to do with the essentials of QM although the story
can be told using the vocabulary and notation of QM.  Likewise, for this case
the mysteries of entanglement are seen as a manifestation of the dependancy of
the statistics on the square of the sum of the fields, a phenomenon unrelated
to QM.

The isomorphism of the mathematics describing spin and polarisation assures us
that with suitable vocabulary, the spin-variant of the EPR gedanken experiment
can also be similarly explained.

\end{document}